\documentclass[12pt]{iopart}

\usepackage{orcidlink}
\usepackage{hyperref}
\hypersetup{
colorlinks=true,
linkcolor=blue,
filecolor=magenta,
urlcolor=blue,
citecolor=blue
}
\usepackage[numbers,sort&compress]{natbib}
\usepackage{amsfonts}

\newcommand{\td}[0]{\mathrm{d}}
\newcommand{\Prob}{\mathcal{P}}

\usepackage{iopams}

\begin{document}

\title{New modeling of the stray light  noise in the main arms of the Einstein Telescope}

\author{M.~Andrés-Carcasona$^1$\orcidlink{0000-0002-8738-1672}, J.~Grandes~Umbert$^1$, D.~González-Lociga$^1$\orcidlink{0009-0009-6991-3960}, M.~Martínez$^{1,2}$\orcidlink{0000-0002-3135-945X}, Ll.~M.~Mir$^1$\orcidlink{0000-0002-4276-715X}, H.~Yamamoto$^3$\orcidlink{0000-0001-6919-9570}}
\address{$^1$Institut de Física d'Altes Energies (IFAE), The Barcelona Institute of Science and Technology, Campus UAB, E-08193 Bellaterra (Barcelona), Spain\\
$^2$Catalan Institution for Research and Advanced Studies (ICREA), E-08010 Barcelona, Spain\\
$^3$LIGO laboratory, California Institute of Technology (Caltech), Pasadena, CA, US}
\ead{mandres@ifae.es}


\begin{abstract}
Stray light represents a significant noise source for gravitational wave detectors, requiring an accurate modeling and mitigation to preserve the experiment's sensitivity.  In this article, we present an updated and improved analysis of the stray-light induced noise in the Einstein Telescope main arm.  The results presented here supersede previous studies taking into account a  number of improvements,  including baffle clipping effects, new numerical calculations for computing diffraction contributions and the influence of baffle edge serrations. Results are presented for both triangular and L-shaped configurations for the experiment. Furthermore, in the case of the triangular configuration,  we examine non-ideal optical cavity conditions, such as beam offsets and misalignments, and the presence of point absorbers in the mirrors, which can increase scattered light noise.  This can be translated into future stringent constraints on the design and operating parameters of the interferometer, thus facilitating to achieve its sensitivity targets. 


\end{abstract}

%
%
%
%
%

\section{Introduction}
Stray light, also known as scattered light, is a well-known noise source in gravitational wave (GW) detectors that can significantly limit their sensitivity if not effectively monitored and mitigated \cite{Accadia:2010zzb,Fiori:2020arj,Was:2020ziy,LIGO:2020zwl,Longo:2020onu,Longo:2021avq,Longo:2023vac,Andres-Carcasona:2022imx,Andres-Carcasona:2024hel,Macquet:2022simsVirgo,Romero-Rodriguez:2020bys,Romero-Rodriguez:2022mje,Ballester:2021bua,Bianchi:2021unp}. This phenomenon arises when photons deviate from their intended paths, scattering off surfaces and ultimately recombining with the main beam \cite{Vinet96,Vinet97,Brisson98,Thorne89,FlanaganThorne95_Diff,Flanagan94,Flanagan95}. Such recombination can introduce phase changes indistinguishable from genuine GW signals, thereby compromising the experimental sensitivity. Current ground-based detectors employ rigorous strategies to mitigate this noise by intercepting and absorbing stray photons. A common solution involves installing baffles, which are precisely shaped, optically coated surfaces made from materials with high optical absorption. These baffles are strategically positioned along the beam tube or near sensitive optics to capture stray light. Their design minimizes surface reflections and scattering, ensuring intercepted light is either absorbed or redirected away from critical optical components \cite{Brisson98,Thorne89,FlanaganThorne95_Diff}. The next generation of ground-based GW detectors, the Einstein Telescope (ET) \cite{ETcds,ETdesign} and the Cosmic Explorer (CE) \cite{CE1,CE2,CE3}, will require similar methods in order to ensure that this noise source will not pose a problem for the scientific outcome~\cite{Abac:2025saz}. 
In this paper we focus on the aspects directly related to ET.

The ET experiment, in its original design, is  envisaged as an underground $\Delta$-shaped experiment with 10~km of arm length, formed by two sets of three nested interferometers optimized at different frequency ranges. In total,  ET will be constituted by six independent Fabry-P\'erot (FP) resonators.   The low-frequency (ET-LF) triangular configuration uses a 1550~nm laser,  includes cryogenics for cooled mirrors operating at 10 K, and stores  about 18~kW of laser power in the FP optical cavities.  The high-frequency (ET-HF) triangular configuration uses a laser of 1064~nm, operates at room temperature,  and accumulates about 3~MW of power in the optical cavities.  Therefore, their optical characteristics are rather different and the studies are performed separately for each configuration.   A preliminary study on scattered light in ET, including an initial baffle layout and design, was presented in Ref.~\cite{Andres-Carcasona:2023qom}. For the proposed baffle layout, the scattered light noise was estimated and shown to be subdominant within the relevant frequency range. The ET collaboration  is now considering an alternative distributed geometry for the experiment with two L-shaped underground interferometer with 15~km of arm length, including both HF and LF technologies.  Therefore, in this paper we also provide scattered light noise estimations for the L-shaped detector configurations.  

The results in this paper build upon the previous study in Ref.~\cite{Andres-Carcasona:2023qom}. In particular, we adopted the same approach to determine the baffle apertures and the baffle layout  along the optical cavity. We enhance several aspects of the modeling, leading to a more accurate noise estimation inside the cavity. The key improvements introduced in this work are:
\begin{itemize}
    \item {\it{Inclusion of the baffles in the simulation}}. In Ref.~\cite{Andres-Carcasona:2023qom}, the effect of baffles on the field was not directly simulated. Instead, the field was computed at the location where a baffle would be placed, with the shielding effect of preceding baffles accounted for analytically via the solid angle viewed from the mirror. In this paper, baffles are explicitly included in the simulation, allowing for a realistic clipping of the field and introducing the effect of diffraction, as additional power can reach the baffles due to this phenomenon.
    \item {\it{Serration of the baffles}}. Baffles are no longer assumed to have smooth edges; instead, serrated edges are considered. This alteration affects the diffraction pattern across subsequent baffles, providing a more realistic simulation.
    \item {\it{Numerical calculation of diffraction noise}}. In the previous study, diffraction noise arising from the finite aperture of baffles was computed analytically based on the work of Ref.~\cite{FlanaganThorne95_Diff}. We introduce a numerical approach to calculate diffraction noise, relaxing the assumptions inherent in the analytical expressions. 
\end{itemize}

\noindent
In addition to an improved implementation of the scattered light simulated noise, in this paper we consider, for the first time, the effects due to offsets and tilts of the laser beam inside the FP cavity. A proper understanding of the tolerances to beam misaslignments for a given baffle layout is instrumental in determining the baseline ET optical configuration.  We determine to which extent the proposed baffle layout can accommodate beam shifts. This can be translated into upper limits on non-ideal laser beam parameters to be considered during the design and operation of the experiment. 

The paper is organized as follows. Section~\ref{sec:modeling} describes in detail the improvements in the simulation of the different noise components. Section~\ref{sec:results} presents the new results for both $\Delta$-shaped and L-shaped  configurations assuming ideal laser beam conditions.  Section~\ref{sec:nonideal}  discusses the effects due to laser beam shifts and point absorbers in the mirrors. Finally, Sec.~\ref{sec:sum} is devoted to conclusions. 



\section{Numerical modeling}\label{sec:modeling}
As pointed out, an accurate modeling of the scattered-light noise in the interferometer arms and the resulting impact on the sensitivity of the experiment to GWs is essential for effectively designing baffles and ensuring a proper noise mitigation strategy is put in place. This section outlines the significant changes in the numerical calculations of scattered light compared to the preliminary work presented in Ref.~\cite{Andres-Carcasona:2023qom}. These changes have been made possible thanks to the new version of the FFT-based simulation code \emph{Stationary Interferometer Simulations} (SIS) \cite{Romero21}. Each of the improvements and the key aspects of the calculation for each noise component are detailed in the following subsections.

\subsection{Inclusion of baffles in the simulation}
\begin{figure}[htb]
    \centering
    \includegraphics[width=0.85\textwidth]{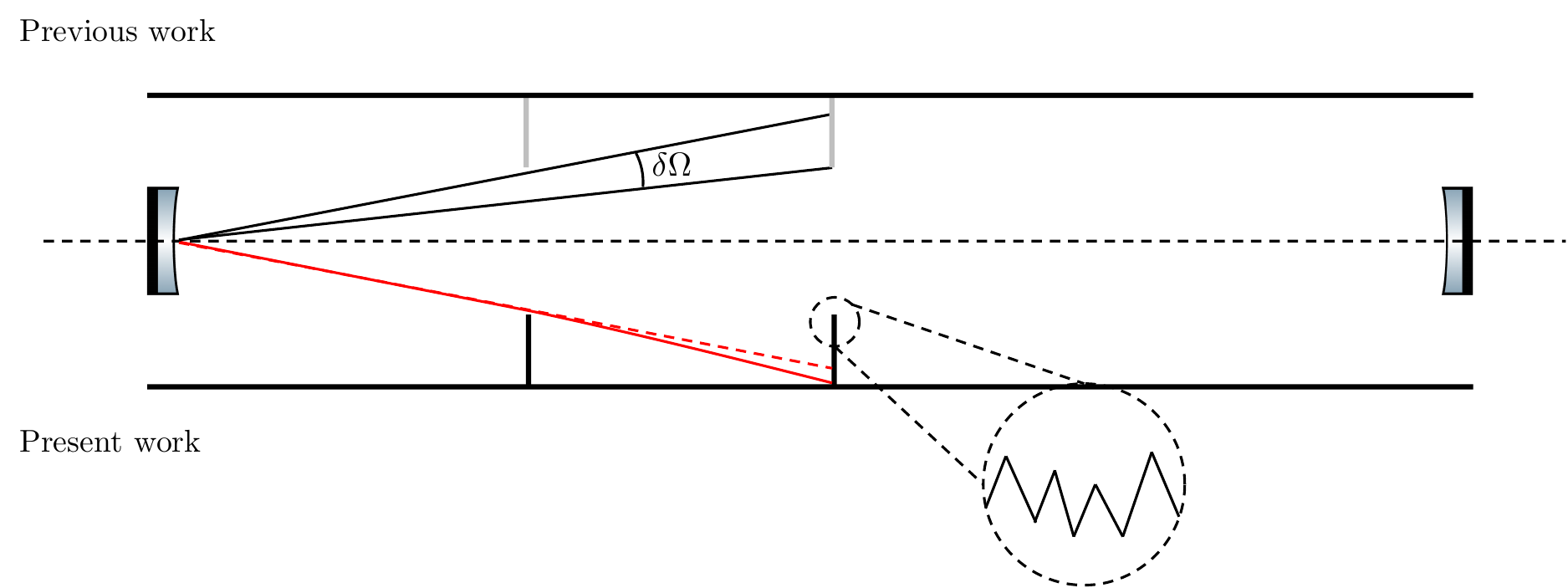}
    \caption{Comparison between the treatment in Ref.~\cite{Andres-Carcasona:2023qom} (upper half) and the approach adopted in this work (lower half) to simulate the baffles and baffle edges. The red dashed line represents the ray-optics trajectory, while the solid red line illustrates how diffraction enhances the light reaching the baffle.}
    \label{fig:scheme}
\end{figure}

In the results presented in Ref.~\cite{Andres-Carcasona:2023qom}, the influence of baffles on the optical field was not explicitly modeled. The cavity was instead treated as if it contained no baffles, and the field was evaluated at the nominal baffle positions while neglecting the shielding provided by upstream baffles. In that framework, the shielding effect was approximated during the noise estimation by incorporating the solid angle~\footnote{Here we consider a cartesian right-handed coordinate system with origin in the input mirror of the optical cavity and the $z$-axis along the laser beam line. The azimuthal angle $\varphi$ is measured around the beam axis and the polar angle $\theta$ is measured with respect to the $z$-axis.}, $\delta \Omega$, subtended by the baffle as viewed from the mirror. This approach is depicted in the upper half of Fig.~\ref{fig:scheme}. Although $\delta \Omega$, which is determined by the design and placement of the baffle, serves as a first-order approximation of the shielding effect, it does not account for the diffraction that occurs in the aperture of the previous baffle. In practice, this diffraction can channel additional light toward subsequent baffles, leading to a higher incident power than predicted by the simplified model. By explicitly including the baffles in our simulation, we obtain a more accurate determination of the power impinging on each baffle, which is essential for a reliable assessment of its contribution to the noise.

\noindent
The fact that baffles are serrated introduces 
 further complexity. Baffles not only clip the field via the inner aperture, but also produce an azimuthally asymmetric diffraction pattern. In the previous work, the analysis assumed that the serration effects were either negligible or could be analytically accounted for. However, the serrations impart a distinct diffraction signature on the field. Accounting for this effect is 
 important for an accurate modeling of the light propagation and its consequent impact on the noise.

\subsection{Diffraction noise}
Diffraction noise occurs when light is clipped by the edges of baffles. The diffracted light can propagate through the cavity and recombine with the main beam, thereby introducing noise. This phenomenon is not confined to individual baffles but can involve coherent interactions among many of them, resulting in cumulative effects. Consequently, the magnitude of this noise increases with the number of baffles present inside the cavity. 
In Ref.~\cite{Andres-Carcasona:2023qom}, the diffraction noise was quantified using the analytical relations derived in Ref.~\cite{FlanaganThorne95_Diff}. With the improvements introduced in SIS, this effect can now be computed numerically, eliminating the assumptions required to derive the analytical relations. The process for each baffle follows these steps:
\begin{itemize}
    \item Scattered light is generated at the input test mass (ITM)  due to its finite aperture and/or surface imperfections.
    \item This field is propagated from the ITM to the position of the baffle.
    \item The field is clipped according to the baffle’s inner aperture and serration pattern.
    \item The clipped field is propagated from the baffle to the end test mass (ETM).
    \item The field is recombined with the main beam.
\end{itemize}

When calculated analytically, as in Ref.~\cite{FlanaganThorne95_Diff}, the diffraction noise is expressed as:

\begin{equation}\label{eq:FlanaganThorne_Diff}
   \fl {h}_{\mathrm{df}}(t) =  \sum_{n=1}^{N_B} \frac{\lambda}{ \pi L} 
\mathbb{I} \bigg\{
\int_0^{2\pi} 
\left[ \frac{\sqrt{\alpha}}{Y_n} f_{\mathrm{sm}}(\pmb{Y}_n) e^{i k Y_n^2 / 2 l_n} \right] \left[ \frac{\sqrt{\alpha}}{Y_n} f_{\mathrm{rm}}(\pmb{Y}_n) e^{i k Y_n^2 / 2 l'_n} \right] 
R X(t) \, \td \varphi \bigg\},
\end{equation}
where $\lambda$ is the wavelength of the laser, $X(t)$ is the baffle displacement in time domain, $k=2\pi/\lambda$ is the wavenumber, $L$ the length of the arms, and $R$ is the beam tube radius. Here, $\mathbb{I}$ represents the imaginary part. The term $\alpha$ is related to the probability of a photon being scattered into an angle $\theta$ as $\alpha/\theta^2$, $\pmb{Y}_n(\varphi)$ is the transverse vector (in the plane of the baffle) from the beam axis to the baffle point at the azimuthal angle $\varphi$, $Y_n=|\pmb{Y}_n(\varphi)|$ and $l_n$ and $l_n'$ represent the distances from the baffle to the ITM and ETM, respectively~\cite{FlanaganThorne95_Diff}. Of particular importance is the structure of the terms within the brackets. The first term in brackets represents the scattered light field that arrives at the edge of the $n^{th}$ baffle, at a given angle $\varphi$ around the baffle. The exponential term is a propagator, and $f_{\mathrm{sm}}$ is a complex factor of order one that quantifies the random variations in phase and amplitude of the light field due to the mirror imperfections. The second term in brackets describes the field after it has propagated to the ETM, interacted with its surface, and scattered back into the main beam. This formulation relies on the reciprocity relation, which establishes a connection between the scattering of light into and out of the main beam~\footnote{See Appendix~B of Ref.~\cite{Flanagan94} for a detailed explanation.}.

Including the effect of radiation pressure, as discussed in Refs.\cite{Andres-Carcasona:2023qom,Hiro_ETMripple}, and leaving the fields inside the integral completely generic, Eq.~(\ref{eq:FlanaganThorne_Diff}) can be reformulated in the frequency domain as:
\begin{equation}
    \tilde{h}_{\mathrm{df}}(f) = \sqrt{\lambda^2+\left(\frac{8 \Gamma P_{\mathrm{circ}}}{cM\pi f^2}\right)^{2} } \frac{X(f)}{\pi L} C\, ,
\end{equation}
being
\begin{equation}
C = \sum_{i=1}^{N_B}\mathbb{I}\left\{\int_{S_{\mathrm{baffle}}} E_{ITM\to B_i} E_{B_i\to ETM} \, \td S \right\}\, ,
\end{equation}
where $E_{ITM\to B_i}$ represents the propagated field from the ITM to the $i^{th}$ baffle, $E_{B_i\to ETM}$ the propagated field from the $i^{th}$ baffle to the ETM, $\Gamma$ the gain of the cavity formed by the ITM and the signal recycling mirror (SRM),  $P_{\mathrm{circ}}$ the circulating power inside the cavity, $M$ the mass of the mirrors, $c$ the speed of light, and $X(f)$ the baffle displacement in the frequency domain and upconverted. The latter is discussed in detail in Sec.~\ref{sec:ground_motion}. The gain $\Gamma$ can be estimated as \cite{Hiro_ETMripple}
\begin{equation}
    \Gamma = \frac{1-r_{ITM}r_{SRM}}{1-r_{ITM}r_{SRM}-r_{ITM}+r_{SRM}}~,
\end{equation}
being $r_{ITM}$ the reflectivity of the ITM and $r_{SRM}$ the one from the SRM.

By simulating these fields inside the integral, SIS does not use any of the analytical relations to propagate them or account for the mirror imperfections and all five processes are explicitly calculated instead of simply approximated. This creates a more reliable estimation of the diffraction noise.

\subsection{Backscatter noise} \label{sec:backscatter_noise}
Baffles are coated with materials designed to maximize light absorption, but some scattering inevitably occurs. This effect can be worsened by imperfections in the coating or surface finish. Even diffuse scattered light can contain angular components that redirect it back towards any of the mirrors and potentially recombine with the main beam. The noise impact becomes more pronounced if the baffles are misaligned or if beam offsets are present.
To account for the backscattering induced noise we use the formalism described in Refs.~ \cite{Thorne89,Flanagan94,Flanagan95,FlanaganThorne95_Diff,Brisson98,Vinet96}. This states that the backscattered light noise is 
\begin{equation}
    \tilde{h}^2_{\mathrm{bs}}(f)=\frac{1}{L^2}\left[\lambda^2+\left(\frac{8 \Gamma P_{\mathrm{circ}}}{cM\pi f^2}\right)^{2}\right]\frac{\td \Prob}{\td \Omega_{bs}} X^2(f) K \, ,
\end{equation}
\noindent
where $\td \Prob / \td \Omega_{bs}$ denotes the probability of a photon getting scattered by the baffle per solid angle  and
\begin{equation}\label{eq:K}
 K = \sum_i K_i =\sum_i 
    \frac{1}{z_i^2}\left( \frac{\td \Prob_i}{\td \Omega_{ms}}\right)^2\delta \Omega_{ms}^i\, ,
\end{equation}
\noindent
being $z_i$ the distance between the ITM and the $i$th baffle, 
$\delta \Omega_{ms}^i$ the solid angle of the $i^{th}$ baffle as seen by the photon being scattered off the mirror and $\td \Prob_i/\td \Omega_{ms}$ the probability of a photon of being scattered by the mirror in the direction of the baffle. This probability can be estimated from the simulation output by integrating the amount of power that deposits into a baffle. This is 
\begin{equation}
    \frac{\td \Prob_i}{\td \Omega_{ms}} \approx  \frac{P_i/P_{\mathrm{circ}}}{\delta \Omega_{ms}^i}~,
\end{equation}
\noindent
where  $P_i$ is the power that reaches the $i^{th}$ baffle, which is estimated from the field, $E$, as
\begin{equation}
    P_i\approx \sum_{k}\sum_{j} |E_{jk}|^2 A_{jk}~,
\end{equation}
\noindent
where $A_{jk}$ is the baffle area illuminated at each cell of the grid in SIS.

The amount of scattered light that a mirror generates is closely tied to the microscopic deviations of the surface when compared to the ideal one. These surface irregularities, or roughness, are generally described by the so-called mirror map, a matrix containing the height deviation of the surface for each point.  The scattering behavior by any surface is generally described by the bidirectional reflectance distribution function (BRDF), which represents the ratio between the differential surface radiance and the surface irradiance. The BRDF is related to the differential power scattered into a solid angle $\td \Omega_s$ relative to the incident power $P_i$, as~\cite{stover2012optical}
\begin{equation}
    \mathrm{BRDF} = \frac{1}{P_i\cos \theta_s}\frac{\td P_s}{\td \Omega_s} = \frac{1}{\cos\theta_s}\frac{\td \Prob_s}{\td \Omega_s}\, ,
\end{equation}
\noindent
where the subscripts $s$ and $i$, refer to the scattered and incident quantities, respectively. The mirror map is linked to the BRDF as \cite{stover2012optical}
\begin{equation} \label{eq:BRDFdef}
    \mathrm{BRDF} = \frac{16\pi^2}{\lambda^4}\cos (\theta_i) \cos (\theta_s) Q S(f_x,f_y)\, ,
\end{equation}
where $Q$ represents the geometrical mean of the specular reflectances measured at the incident and scattered angles and $S(f_x,f_y)$ is the 2-dimensional power spectral density (PSD) of the surface \cite{HiroPSD}. In this work, we use the mirror maps projected for Virgo in O5 as depicted in Refs.~\cite{Andres-Carcasona:2023qom,Romero-Rodriguez:2022mje, Macquet:2022simsVirgo}.

In the study presented in Ref.~\cite{Andres-Carcasona:2023qom}, it was discussed how the first half set of baffles could not be simulated due to the limitations of the mirror map. This is caused by its finite resolution, which sets a limit to the maximum scattering angle that can be resolved. This problem was overcome by simulating the far half of the tube and then doubling the result. In this work, we use instead the new capability of SIS to include the BRDF tail~\cite{Romero21}, so that for angles larger than this limit, the field is estimated from experimental measurements of the BRDF or employing an analytical model. In this way, the calculation for all baffles can be performed using SIS. 

\subsection{Ground motion} \label{sec:ground_motion}
The ground motion is a fundamental input to the models of scattered light noise, as it provides an external modulation of the field that couples into the interferometer resonating field. In particular, low-frequency seismic motion, which is dominant below a few Hz, can be upconverted into the detector's most sensitive band via a highly non-linear processes~\cite{Ottaway:2012oce,Chua:2013fgl,Was:2020ziy}. This effect is known as phase-wrapping. Therefore, a realistic and conservative estimation of the baffles' motion must be used  to reliably estimate the potential contribution of scattered light to the total noise budget.

\begin{figure}[htb]
    \centering
    \includegraphics[width=0.85\textwidth]{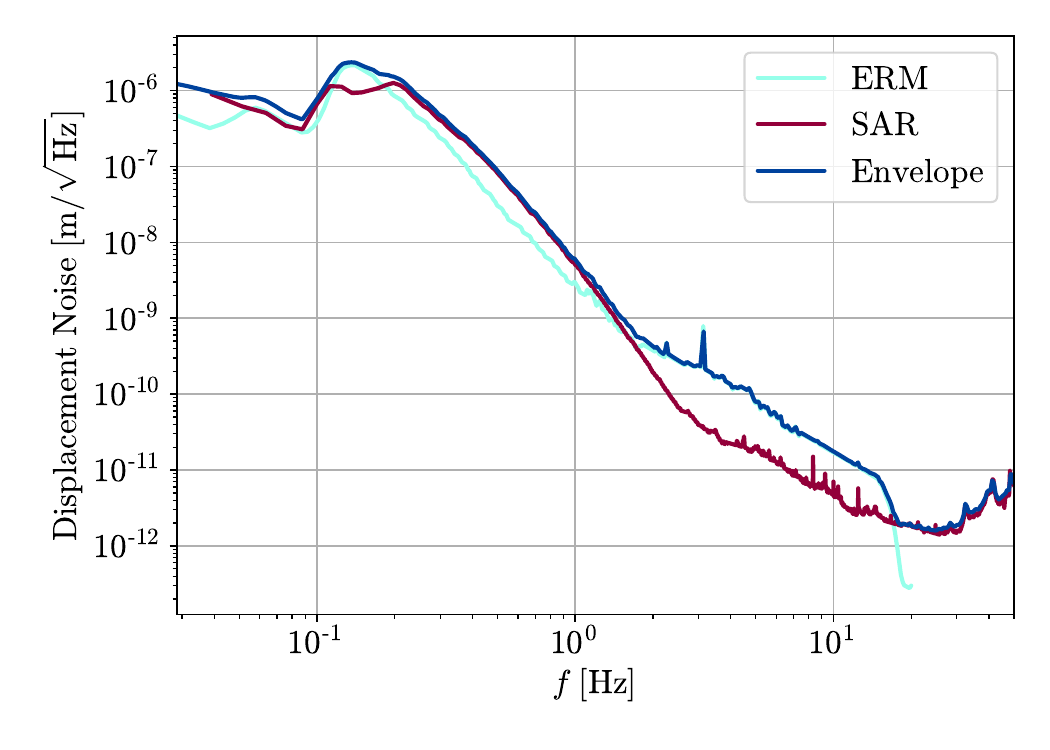}
    \caption{Measured underground seismic motion as a function of frequency at two ET candidate sites, Euregio Rhein-Maas (ERM) and Sardegna (SAR), and the resulting envelope (see text).}
    \label{fig:NoiseEnvelope}
\end{figure}

As it was the case in Ref.~\cite{Andres-Carcasona:2023qom}, we considered two possible locations:  the Sos Enattos mine area in Sardegna~\cite{Naticchioni:2014bra,SoSEnattos1,SoSEnattos2,allocca2021seismic,saccorotti2023array,di2021seismological,di2023temporal} in Italy, and the Euregio Rhein Maas~\cite{ERM,koley2019characteristics,Bader:2022tdz} area in the borders between Belgium, The Netherlands and Germany. Thanks to thorough campaigns carried out  in both sites to characterize the seismic motion, seismic data are available and will be input to our calculations.
In this paper, we adopt a site-independent approach by constructing an envelope of the measured 90$\%$ confidence level ground motion spectra measured at Sos Enattos and Euregio~\footnote{A third potential candidate site is now being considered in the area of Lausitz (Germany). Although the campaigns of seismic measurements have not been concluded yet, first preliminary results indicate the level of seismic noise at Lausitz would be comparable and within the range of those inputs for our studies.}. This envelope represents a conservative upper bound on the seismic motion, thereby capturing the worst-case scenario for scattered light coupling. By using this envelope rather than a site-specific model, we ensure that our results remain broadly applicable while still respecting realistic bounds derived from in-situ measurements. The original seismic motions for each of the two sites and the envelope are shown in Fig.~\ref{fig:NoiseEnvelope}.

Finally, we have assumed the baffle motion follows exactly that of the ground. Therefore,  we adopt a mechanical transfer factor between the ground motion and the baffle of one, independently of the frequency. This motion is then upconverted using the methodology described in Ref.~\cite{Andres-Carcasona:2023qom}. The use of a more realistic description of the baffle motion is out of the scope of this analysis as it requires a mature implementation of the designs for the vacuum tube, the baffles and the different mechanical supports, which are currently partially unknown. 


\section{Results in nominal conditions}
\label{sec:results}

Using the formalism described in the previous section, the scattered light noise can be estimated.
Relevant optical parameters of the cavities~\footnote{A complete set of parameters of the baffle geometry and apertures is presented in Ref.~\cite{Andres-Carcasona:2023qom}.} for both 
$\Delta$-shaped and L-shaped configurations are collected in Tables~\ref{tab:GeneralParams_10km} and \ref{tab:GeneralParams_15km}, respectively, and ideal laser beams, perfectly centered in the cavity,  are assumed. The simulated noise results depend on the baffle design and optical performance, as well as on the separation between baffles and the total number of baffles installed in the cavity.  As in Ref.~\cite{Andres-Carcasona:2023qom}, baffle apertures are fixed along the tube and are determined by the level of clipping losses close to the mirrors, and the baffles are strategically placed along the vacuum tube such that the tube walls are not exposed to scattered light contributions from the mirrors. In addition, baffles are located at the ends of the 50~m long vacuum tube sections, leading to an asymptotic maximal separation of 50~m between baffles along the vacuum tube, away from the mirrors. Finally, the contribution of baffles in the cryotrap areas, covering the first 12~m (ET-HF) and 52~m (ET-LF) of the vacuum tube in the $z$ direction are not included in the calculations, as they are the subject of a dedicated study.

\renewcommand\arraystretch{1.1}
\begin{table}[htb]
\begin{center}
\footnotesize
\begin{tabular}{c |c|c |c|l}
\hline \hline
\multicolumn{5}{c}{Cavity parameters for the $\Delta$-shaped ET with $10$ km arms} \\ \hline 
{Variable} & {ET-HF} &{ET-LF} & {Units} & {Description} \\ \hline 
$m$ & $200$ & $211$ & [kg] & Mirror mass \\ \hline 
$L$ & $10$ & $10$ & [km] & Length of an arm \\ \hline 
$\lambda$ & $1064$ & $1550$ & [nm] & Wavelength of the laser \\ \hline
$R_m$ & $0.31$ & $0.225$ & [m] & Radii of the mirrors \\ \hline
$\mathcal{R}_1$  & $5070$ & $5580$ & [m] & Radius of curvature ITM\\ \hline
$\mathcal{R}_2$  & $5070$ & $5580$ & [m] & Radius of curvature ETM\\ \hline
$P_{\mathrm{circ}}$ & $3000$ & $18$ & [kW] & Circulating power in the cavity \\ \hline
$R$  & $0.5$ & $0.5$ &[m] & Radius of the vacuum pipe. \\ \hline 
RTL & $80$ & $45$ & ppm & Round-trip losses \\ \hline
$r_{ITM}$ & $0.99648$ & $0.99648$ & -- & Reflectivity of the ITM mirror \\ \hline 
$r_{SRM}$ & $0.97417$ & $0.89387$ & -- & Reflectivity of the SRM \\ \hline 
$\displaystyle \frac{\td \Prob}{\td \Omega_{bs}}$ & $10^{-4}$ & $10^{-4}$& [str$^{-1}$] & BRDF  of the baffles \\ \hline 
$z_0$ & $12.35$ & $52.35$ & [m] & Position of the first baffle \\ \hline
$N$ & $270$ & $228$ & --& Number of baffles per arm \\ \hline \hline
\end{tabular}
\end{center}
\caption{Parameters of the FP arm cavities used for the $\Delta$-shaped ET with $10$ km arms. 
The values are extracted from Refs.~\cite{ETcds,ETdesign,ET_NoiseCurve,Andres-Carcasona:2023qom}.}
\label{tab:GeneralParams_10km}
\end{table}

\renewcommand\arraystretch{1.1}
\begin{table}[htb]
\begin{center}
\footnotesize
\begin{tabular}{c |c|c |c|l}
\hline \hline
\multicolumn{5}{c}{Cavity parameters for the L-shaped ET with $15$ km arms} \\ \hline 
{Variable} & {ET-HF} &{ET-LF} & {Units} & {Description} \\ \hline 
$m$ & $200$ & $211$ & [kg] & Mirror mass \\ \hline 
$L$ & $15$ & $15$ & [km] & Length of an arm \\ \hline 
$\lambda$ & $1064$ & $1550$ & [nm] & Wavelength of the laser \\ \hline
$R_m$ & $0.31$ & $0.225$ & [m] & Radii of the mirrors \\ \hline
$\mathcal{R}_1$  & $7745$ & $10697$ & [m] & Radius of curvature ITM\\ \hline
$\mathcal{R}_2$  & $7745$ & $10697$ & [m] & Radius of curvature ETM\\ \hline
$P_{\mathrm{circ}}$ & $3000$ & $18$ & [kW] & Circulating power in the cavity \\ \hline
$R$  & $0.5$ & $0.5$ &[m] & Radius of the vacuum pipe. \\ \hline 
RTL & $80$ & $45$ & ppm & Round-trip losses \\ \hline
$r_{ITM}$ & $0.99648$ & $0.99648$ & -- & Reflectivity of the ITM mirror \\ \hline 
$r_{SRM}$ & $0.97417$ & $0.89387$ & -- & Reflectivity of the SRM \\ \hline
$\displaystyle \frac{\td \Prob}{\td \Omega_{bs}}$ & $10^{-4}$ & $10^{-4}$& [str$^{-1}$] & BRDF  of the baffles \\ \hline
$z_0$ & $12.35$ & $52.35$ & [m] & Position of the first baffle \\ \hline
$N$ & $370$ & $328$ & --& Number of baffles per arm \\ \hline \hline
\end{tabular}
\end{center}
\caption{Parameters of the FP arm cavities used for the L-shaped ET with $15$ km arms. 
The values are extracted from Refs.~\cite{ETcds,ETdesign,ET_NoiseCurve,Andres-Carcasona:2023qom}.}
\label{tab:GeneralParams_15km}
\end{table}

\begin{figure}[htb]
    \centering
    \includegraphics[width=0.85\textwidth]{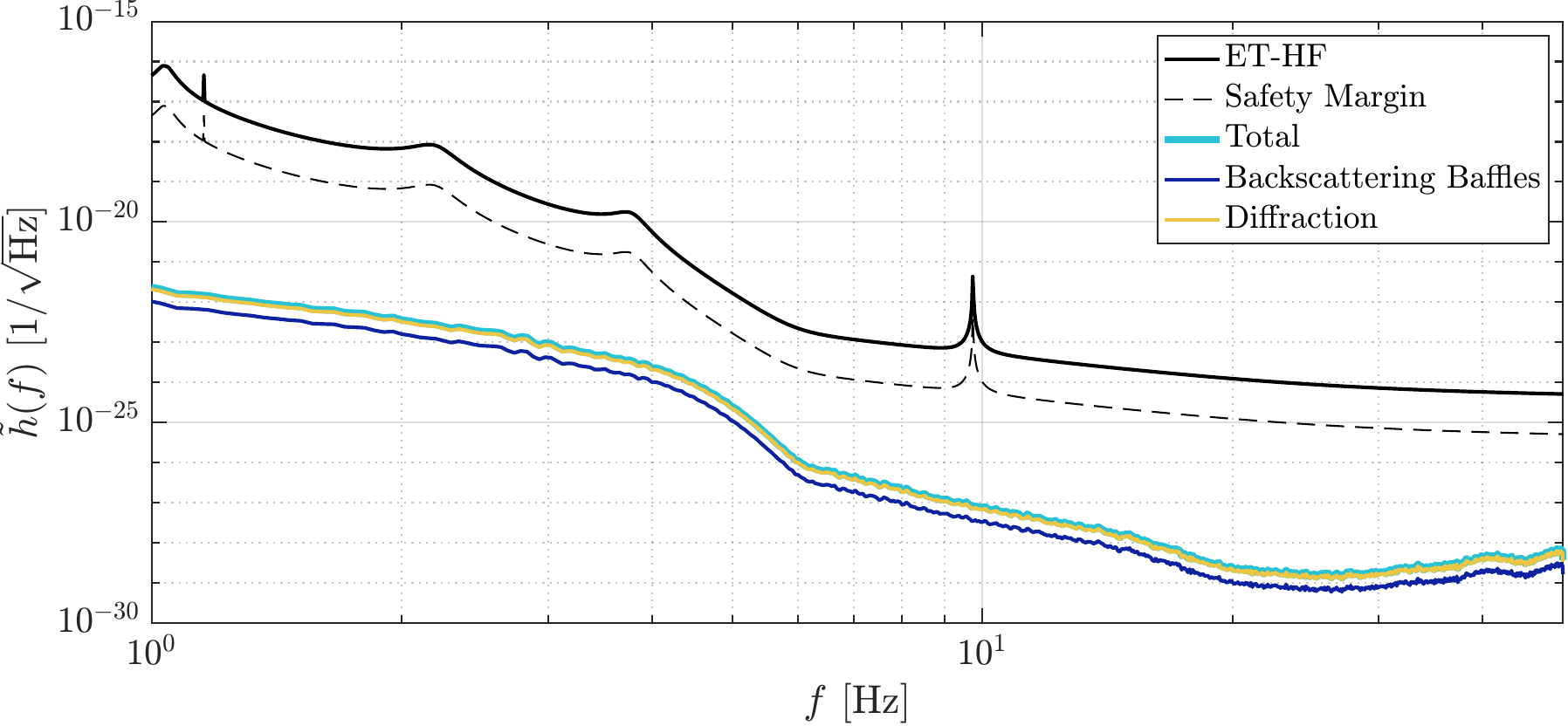}
    \includegraphics[width=0.85\textwidth]{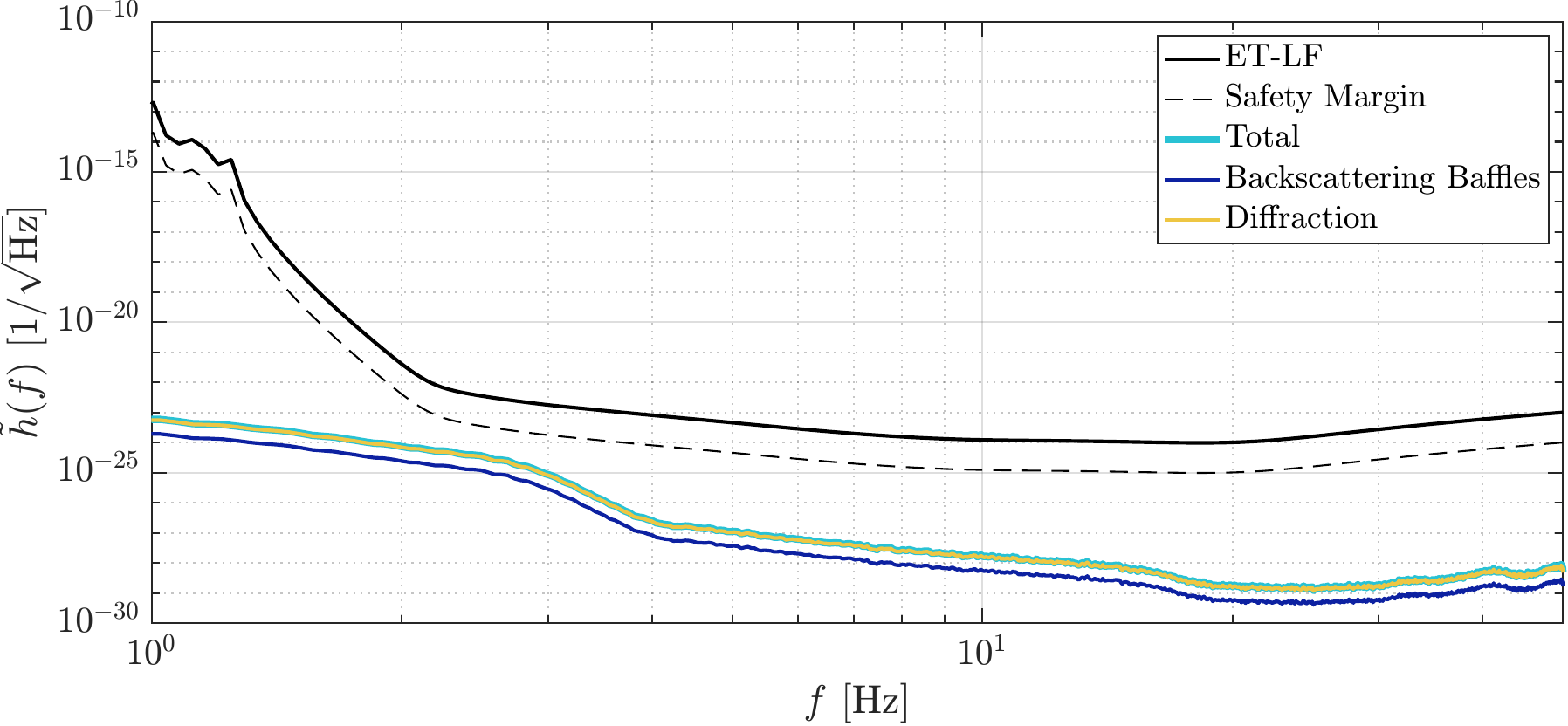}
    \caption{Stray light noise due to diffraction effects (yellow lines), backscattering effects (navy blue lines), and the total noise (cyan
lines) as a function of frequency compared to the anticipated (top) $\Delta$-shaped ET-HF and (bottom) $\Delta$-shaped ET-LF sensitivity curves (black lines)
and the corresponding 1/10 safety margin (dashed lines).}
    \label{fig:NoiseTriangle10km}
\end{figure}

Figure~\ref{fig:NoiseTriangle10km} presents the results for the total scattered light  noise for the $\Delta$-shaped 10~km ET as a function of frequency, compared to the anticipated sensitivity. The frequency range covered is dictated by the minimum frequency for which the ET sensitivity is specified (1 Hz) and by the maximum frequency of the available seismic noise curve (50 Hz).  The results are presented separately for the ET-HF and ET-LF cases.  The noise is dominated by the contribution from diffraction, followed by the backscattering contribution, which is typically a factor  ten  (four) smaller for the ET-HF (ET-LF) configuration. 
Compared to the previous results in Ref.~\cite{Andres-Carcasona:2023qom}, the predicted noise has increased by about 50$\%$ in the case of ET-HF,  and it has decreased by about 30$\%$ for ET-LF, dominated by the  updated diffraction noise calculation. The total induced noise is typically many orders of magnitude smaller than the nominal sensitivity. In the frequency range where the noise levels are closer to the sensitivity curves (2~Hz to 5~Hz) the noise still remains at least two orders of magnitude below. These results indicate that the predicted scattered light noise, with the proposed baffle configuration, does not constitute a limitation for the ET sensitivity.  

\begin{figure}[htb]
    \centering
    \includegraphics[width=0.85\textwidth]{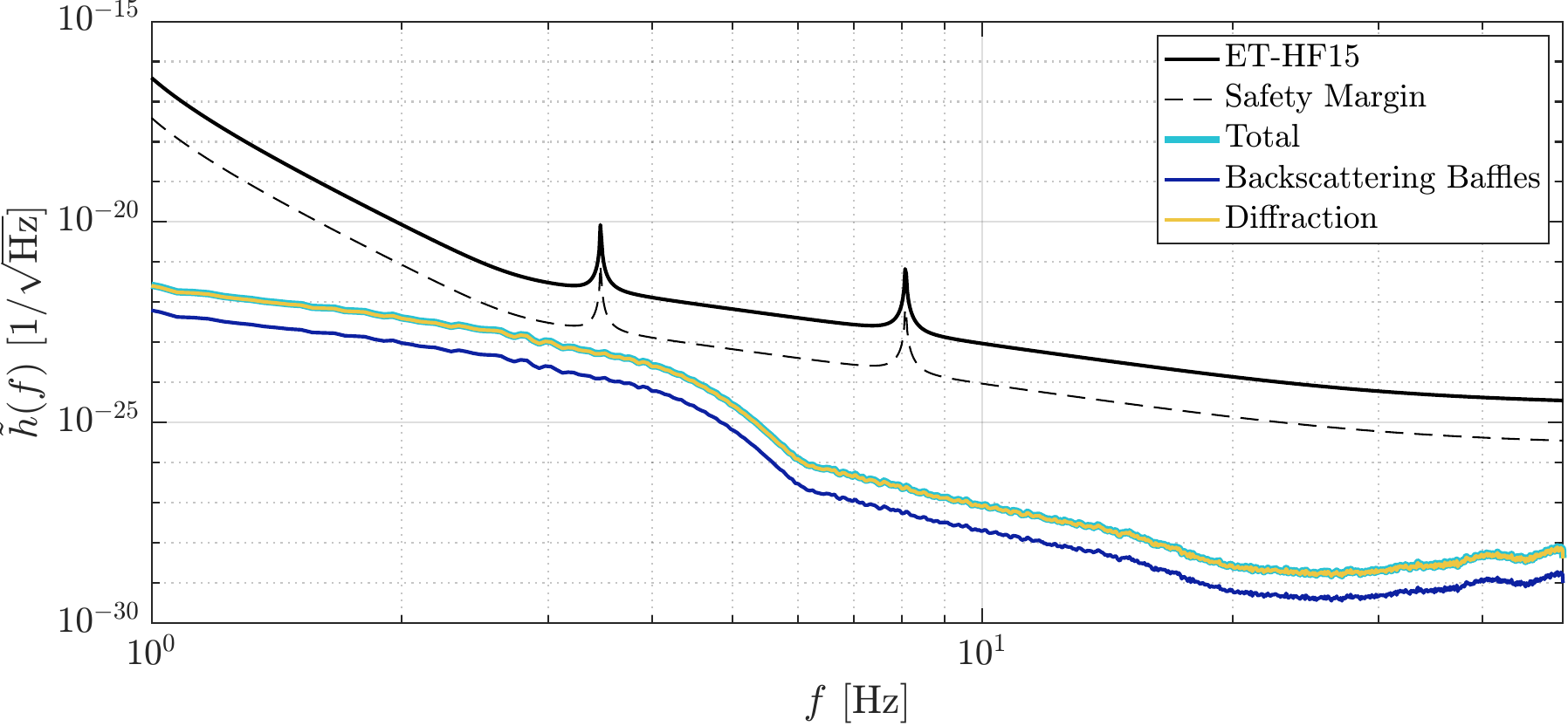}
    \includegraphics[width=0.85\textwidth]{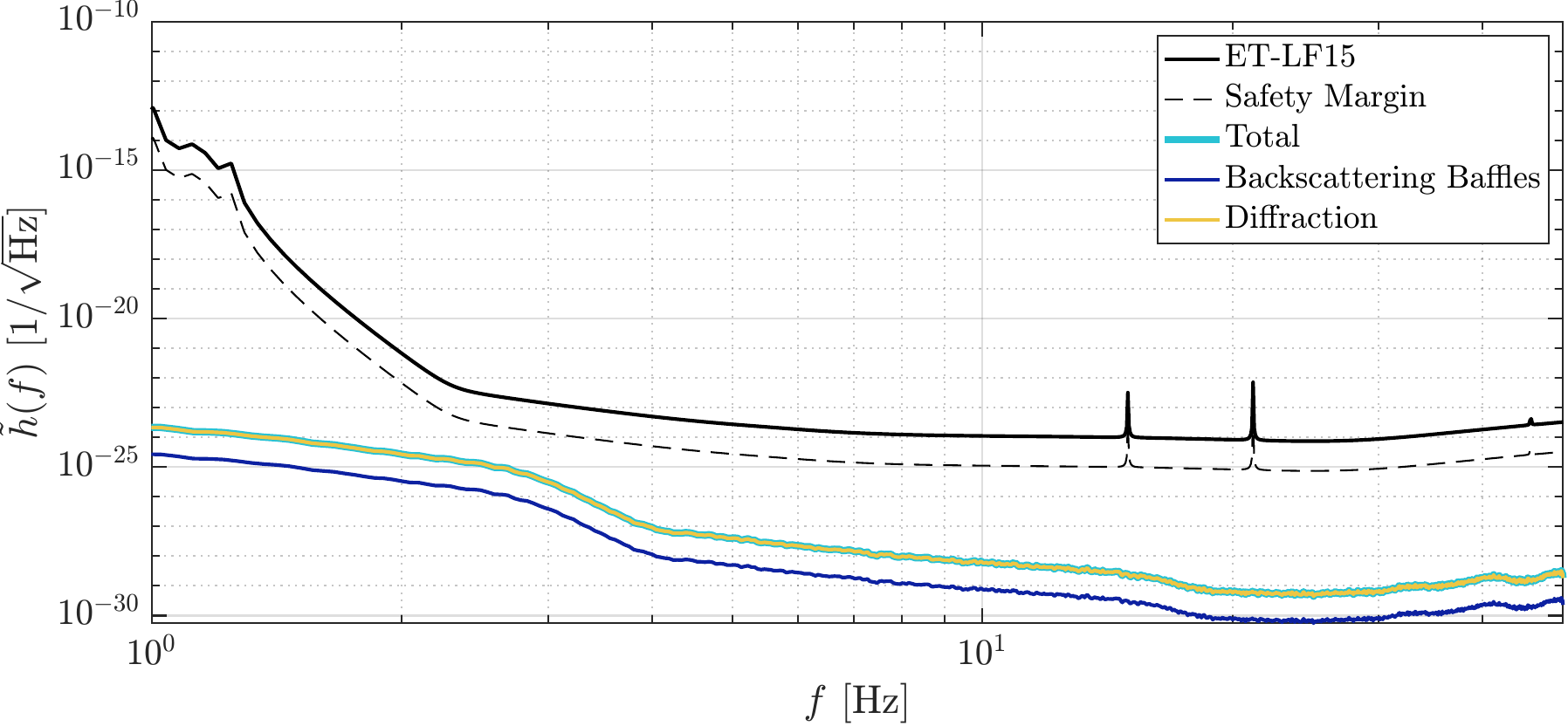}
    \caption{Stray light noise due to diffraction effects (yellow lines), backscattering effects (navy blue lines), and the total noise (cyan lines) as a function of frequency compared to the anticipated (top) L-shaped ET-HF and (bottom) L-shaped ET-LF sensitivity curves (black lines)
and the corresponding 1/10 safety margin (dashed lines).}
    \label{fig:NoiseL15km}
\end{figure}

Figure~\ref{fig:NoiseL15km} shows the results for the L-shaped 15~km ET-HF and ET-LF configurations. Similar conclusions can be drawn, although the noise levels are larger than those in the  $\Delta$-shaped case. This is directly attributed to the fact that a fixed asymptotic separation of 50~m between baffles have been kept along the vacuum tube, translating into a hundred more baffles installed per arm. Both diffraction and backscattering noise contributions increase with the increasing number of baffles. Therefore, there is room for improvement to reduce the noise level by increasing the asymptotic separation between baffles and thus reducing the total number of baffles installed. Nevertheless, such optimization must take into consideration that baffles will be   required to hide from the mirrors the vacuum pumping structures that are foreseen every few hundred meters along  the vacuum pipe.


%
%

\section{Results in non-ideal conditions}\label{sec:nonideal}

The scattered light noise is computed in the presence of offsets and tilts in the laser beam inside the cavity for the $\Delta$-shaped ET-HF and ET-LF configurations.  We also studied the impact of point absorbers in the mirrors.

\subsection{Effect of beam offsets}
\label{sec:offset}

\begin{figure}[htb]
    \centering
    \includegraphics[width=0.85\textwidth]{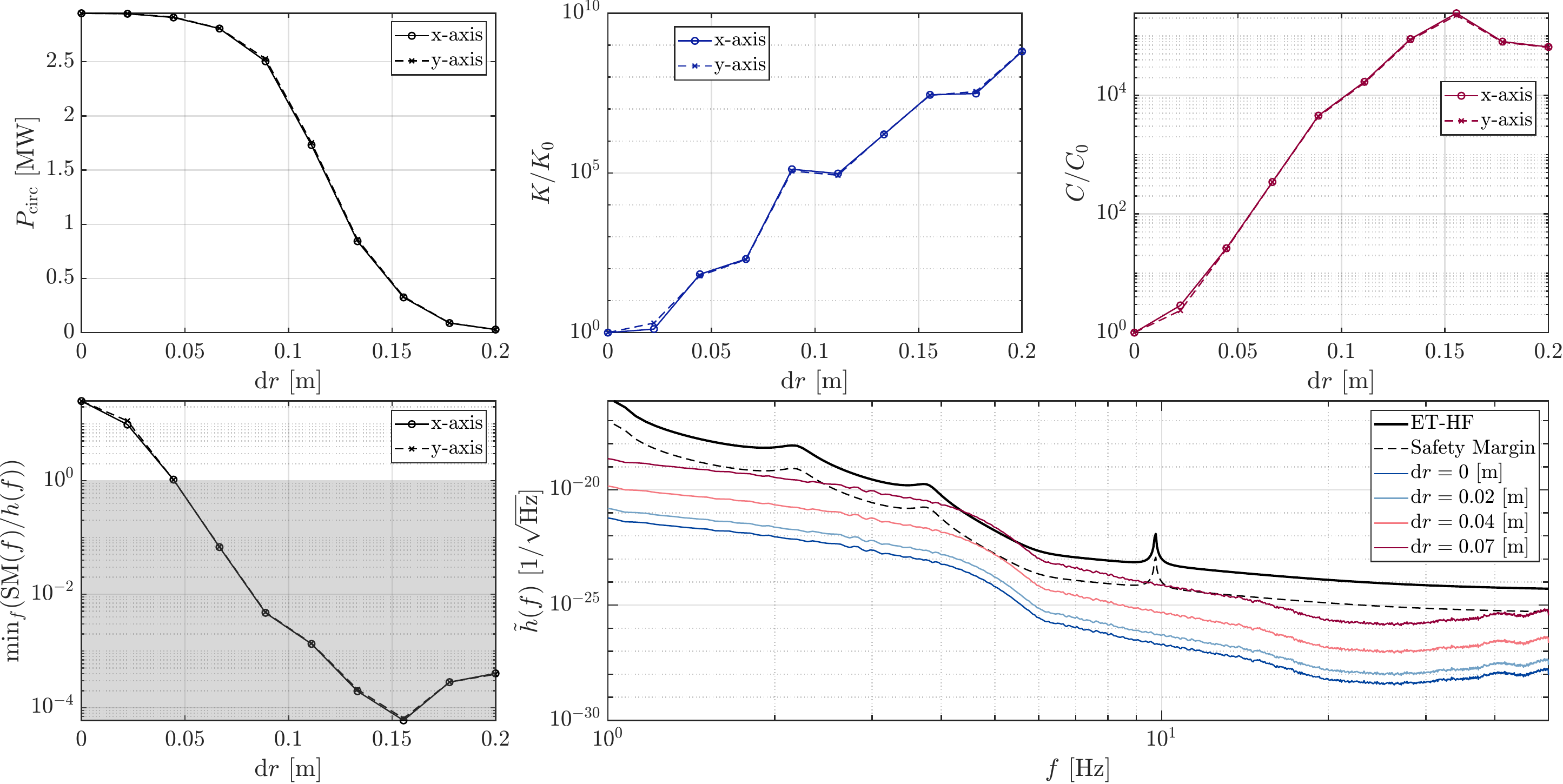}
    \caption{
    (top) For $\Delta$-shaped ET-HF, the circulating power in the cavity as a function of $\td r$, $P_{\rm circ}$; and the evolution of backscattering, $K/K_0$, and diffraction, $C/C_0$, noise terms as a function of $\td r$, where $K_0$ and $C_0$ denote the corresponding values for $\td r = 0$.  Results are presented for displacements along the $x$-axis (solid lines) and the $y$-axis (dashed lines).
    (bottom) For $\Delta$-shaped ET-HF, the resulting ${\rm{min}}_f (\mathrm{SM}(f)/h(f))$ as a function of $\td r$  for  displacements along the $x$-axis (solid lines) and the $y$-axis (dashed lines); and the resulting scattered light noise curve for different  $\td r$ values (colored lines) compared to the ET-HF sensitivity curve (black line) and the corresponding 1/10 safety margin (dashed line).}
    \label{fig:ETHFoffset}
\end{figure}

\begin{figure}[htb]
    \centering
    \includegraphics[width=0.85\textwidth]{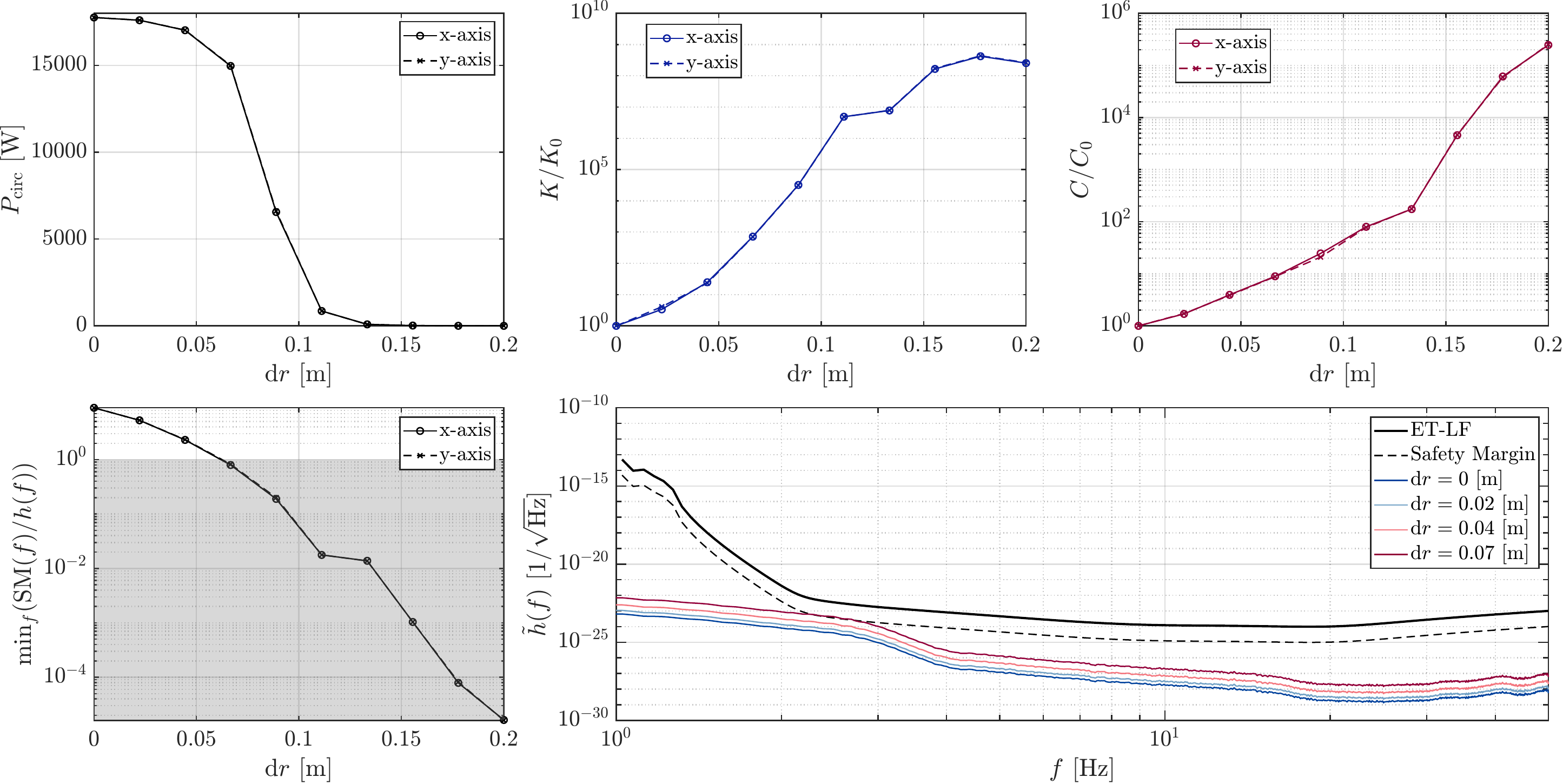}
    \caption{
    (top) For $\Delta$-shaped ET-LF, the circulating power in the cavity as a function of $\td r$, $P_{\rm circ}$; and the evolution of backscattering, $K/K_0$, and diffraction, $C/C_0$, noise terms as a function of $\td r$, where $K_0$ and $C_0$ denote the corresponding values for $\td r = 0$.  Results are presented for displacements along the $x$-axis (solid lines) and the $y$-axis (dashed lines).
    (bottom) For $\Delta$-shaped ET-LF, the resulting ${\rm{min}}_f (\mathrm{SM}(f)/h(f))$ as a function of $\td r$  for  displacements along the $x$-axis (solid lines) and the $y$-axis (dashed lines); and the resulting scattered light noise curve for different  $\td r$ values (colored lines) compared to the ET-LF sensitivity curve (black line) and the corresponding 1/10 safety margin (dashed line).}
     \label{fig:ETLFoffset}
\end{figure}

We first consider offsets in the $x-y$ plane. Radial offsets $\td r = \sqrt{dx^2+dy^2}$ up to 20~cm have been explored. Different azimuthal directions have been considered leading to very similar results. Here,  for illustration,  we present offsets along the $x$ and $y$ axes. The results for ET-HF and ET-LF are collected in Figures~\ref{fig:ETHFoffset} and~\ref{fig:ETLFoffset}, respectively. The circulating power in the cavity decreases rapidly as $\td r$ increases reflecting the fact that the cavity goes out of resonance.  As the laser beam approaches the edges of the baffle aperture we observe a rapid increase in the $K$ and $C$ factors related to the backscattering and diffraction noise contributions, respectively, which change  by orders of magnitude compared to those in ideal laser beam conditions.  Similar behavior is observed for ET-HF and ET-LF. As a result, there is  a significant increase of the total scattered-light induced noise in the presence of significant offsets in the cavity.  In  the case of ET-HF, an offset $\td r \sim 4$~cm already increases the noise such that at frequencies about 5~Hz the noise curve meets the 1/10 of the ET sensitivity safety margin (dashed lines in the figures).  For larger values of $\td r$ the total scattered-light induced noise would compromise the sensitivity of the experiment in a wider frequency range. 
In  the case of ET-LF,  the noise curve meets the 1/10 of the ET sensitivity safety margin at about 3~Hz when $\td r \sim 7$~cm. This is consistent with the different beam profiles for ET-HF and ET-LF configurations (being the ET-LF laser beam size smaller at the edges of the cavity) and the way the baffle apertures were determined in Ref.~\cite{Andres-Carcasona:2023qom}, where an eventual beam offset of 5~cm was already taken into account. 

In order to determine the maximum tolerable offset that would allow operating the interferometer without compromising its sensitivity, we monitor the ratio between the 1/10 of ET sensitivity curve, $\mathrm{SM}(f)$, and the resulting scattered light noise, $h(f)$, at the frequency $f$ that minimizes it, ${\rm{min}}_f (\mathrm{SM}(f)/h(f))$, as a function of $\td r$.  As pointed out, in the case of ET-HF, an offset of 4~cm already becomes a risk, whereas for  ET-LF this occurs for $\td r$ of the order of 7~cm. 

\subsection{Effect of beam misalignments}
\label{sec:tilts}

\begin{figure}[htb]
    \centering
    \includegraphics[width=0.85\textwidth]{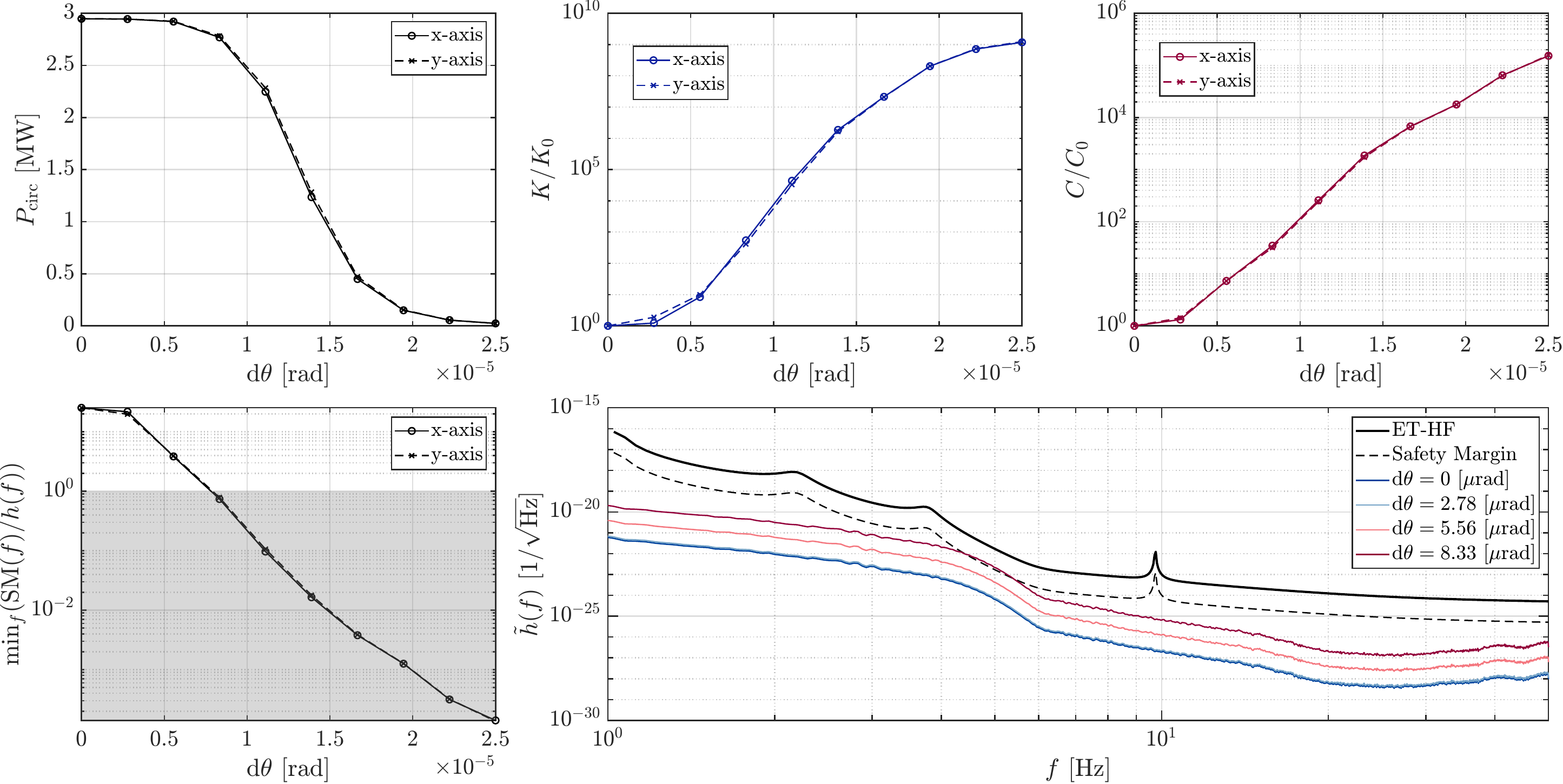}
     \caption{(top) For $\Delta$-shaped ET-HF, the circulating power in the cavity as a function of $\td \theta$, $P_{\rm circ}$; and the evolution of backscattering, $K/K_0$, and diffraction, $C/C_0$, noise terms as a function of $\td \theta$, where $K_0$ and $C_0$ denote the corresponding values for $\td \theta = 0$.  Results are presented for tilts along the $x-z$-plane (solid lines) and the $y-z$ plane (dashed lines).
    (bottom) For $\Delta$-shaped ET-HF, the resulting ${\rm{min}}_f (\mathrm{SM}(f)/h(f))$ as a function of $d\theta$  for  tilts along the $x-z$ plane (solid lines) and the $y-z$ plane (dashed lines); and the resulting scattered light noise curve for different  $\td\theta$ values (colored lines) compared to the ET-HF sensitivity curve (black line) and the corresponding 1/10 safety margin (dashed line).}
    \label{fig:ETHFmiss}
\end{figure}

\begin{figure}[htb]
    \centering
    \includegraphics[width=0.85\textwidth]{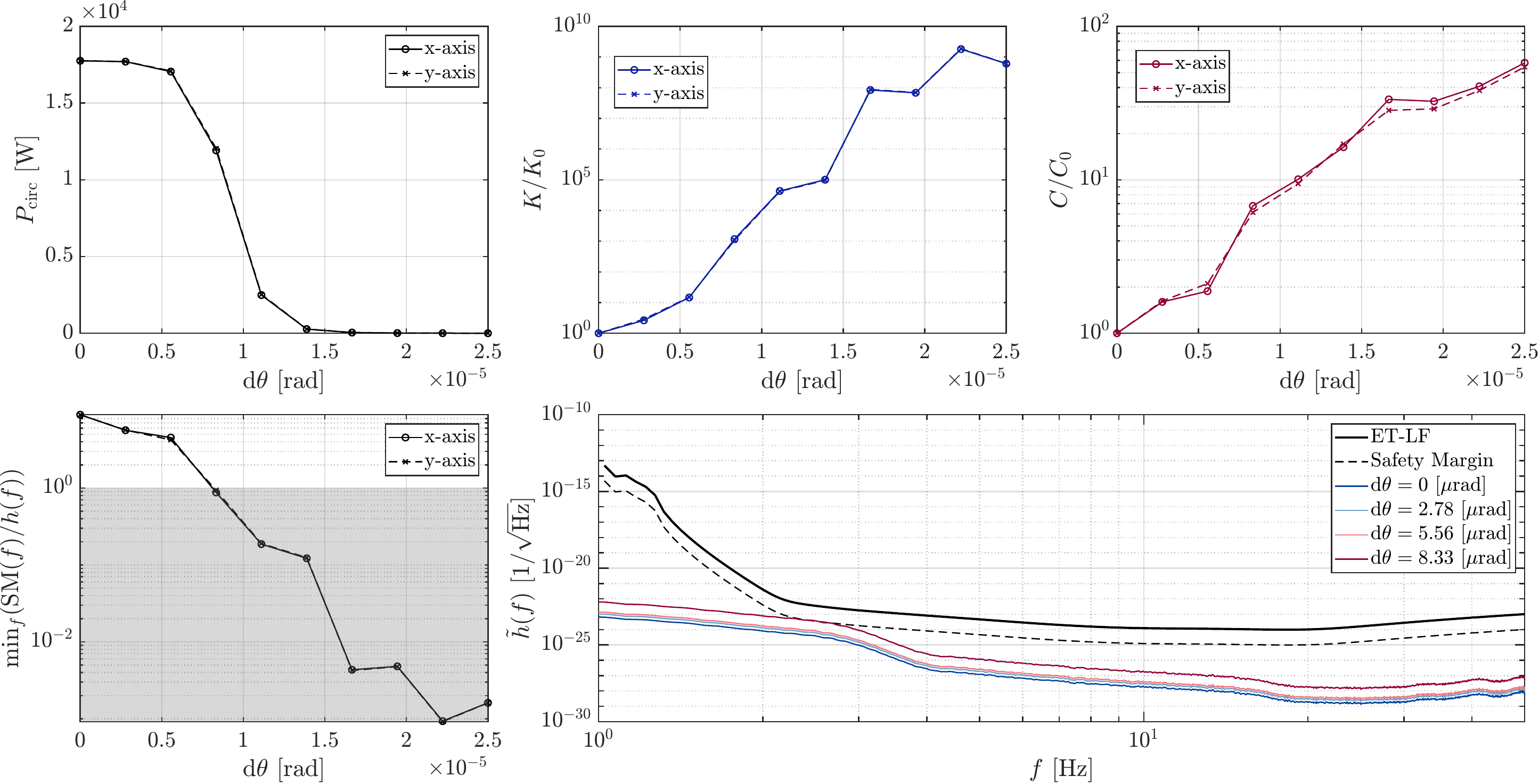}
     \caption{(top) For $\Delta$-shaped ET-LF, the circulating power in the cavity as a function of $\td \theta$, $P_{\rm circ}$; and the evolution of backscattering, $K/K_0$, and diffraction, $C/C_0$ noise terms as a function of $\td \theta$, where $K_0$ and $C_0$ denote the corresponding values for $\td \theta = 0$.  Results are presented for tilts along the $x-z$ plane (solid lines) and the $y-z$ plane (dashed lines).
    (bottom) For $\Delta$-shaped ET-LF, the resulting ${\rm{min}}_f (\mathrm{SM}(f)/h(f))$ as a function of $d\theta$ for  tilts along the $x-z$ plane (solid lines) and the $y-z$ plane (dashed lines); and the resulting scattered light noise curve for different  $\td\theta$ values (colored lines) compared to the ET-LF sensitivity curve (black line) and the corresponding 1/10 safety margin (dashed line).}
    \label{fig:ETLFmiss}
\end{figure}

A study is performed including beam tilts in the cavity, expressed in terms of deviations $\td\theta$ as measured from the ITM center. Values for  $\td\theta$ up to $2.5 \cdot 10^{-5}$~rad have been considered. We explored two possible $\varphi$ directions ($\varphi = 0, \pi$), corresponding to tilts in the $x-z$ and $y-z$ planes and leading to very similar results. Results are collected in Figures~\ref{fig:ETHFmiss} and \ref{fig:ETLFmiss}, for the ET-HF and ET-LF configurations, respectively. As expected, significant tilts translate into a severe decrease of circulating power in the cavity, before going out of resonance, increasing the scattered-light noise contributions.  A similar analysis was performed to determine the maximum tolerable values for the tilt. We conclude that deviations at the level of  $\td\theta \geq 8 \cdot 10^{-6}$~rad would already compromise the sensitivity of the interferometer at $4-6$~Hz.  

%
%

\subsection{Effect of point absorbers}
\label{sec:point}

We studied the impact of point absorbers on ET sensitivity.  We considered the $\Delta$-shaped 10~km ET-HF and ET-LF configurations. For simplicity, only the presence of a single absorber at the ETM is explored. Following the work in the previous section, we monitor the evolution of the cavity circulating power, and the factors governing the backscattering and diffraction noise contributions as a function of the power absorbed in the mirror defect, $P_{\rm{abs}}$, and its position on the mirror surface.  We also monitor the evolution of the scattered light noise curve  compared to the 1/10 ET sensitivity safety margin ${\rm{min}}_f (\mathrm{SM}(f)/h(f))$. Values for $P_{\rm{abs}}$ up to 500~mW have been considered, as inspired by previous studies~\cite{Jia:2021hgr,LIGOScientific:2021kro,aLIGO:2020wna}. The absorbers are placed at different radii, $r$, as computed from the center of the mirror with randomized $\varphi$ positions.  

\begin{figure}[htb]
    \centering
    \includegraphics[width=0.85\textwidth]{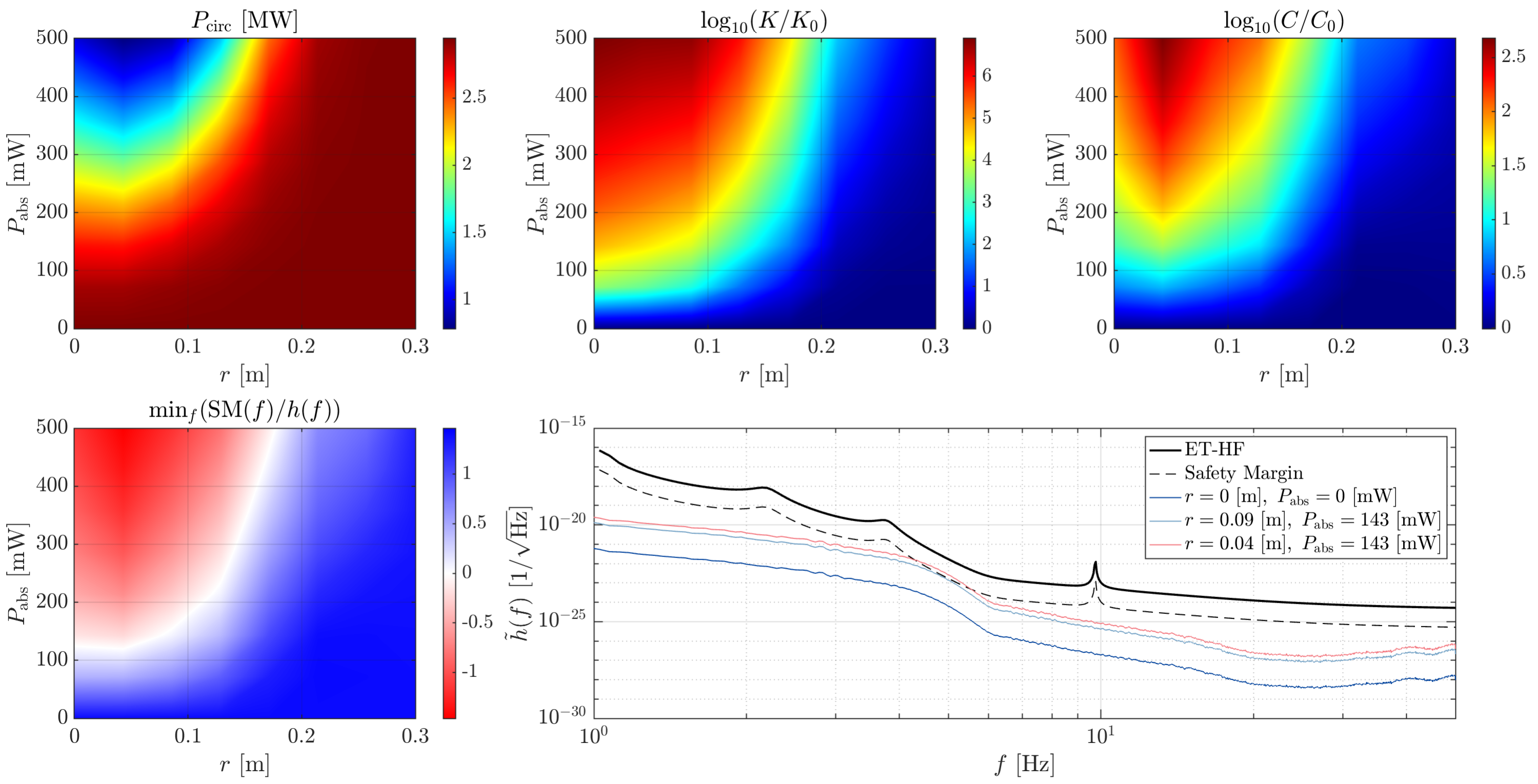}
     \caption{(top) For $\Delta$-shaped ET-HF, the circulating power in the cavity $P_{\rm circ}$ as a function of the position $\td r$ and the power absorption $P_{\mathrm{abs}}$ of the point defect in the mirror; and the evolution of backscattering, $K/K_0$, and diffraction, $C/C_0$, noise terms as a function of $\td r$ and $P_{\mathrm{abs}}$, where $K_0$ and $C_0$ denote the corresponding values for $dr =  0, \ P_{\rm abs} = 0$.
    (bottom) For $\Delta$-shaped ET-HF, the resulting ${\rm{min}}_f (\mathrm{SM}(f)/h(f))$ as a function of $\td r$ and $P_{\rm circ}$,  and the resulting scattered light noise curve for different  $\td r$ and  $P_{\rm abs}$ values compared to the ET-HF sensitivity curve (black line) and the corresponding 1/10 safety margin (dashed line).}
    \label{fig:ETHFabs}
\end{figure}

\begin{figure}[htb]
    \centering
    \includegraphics[width=0.85\textwidth]{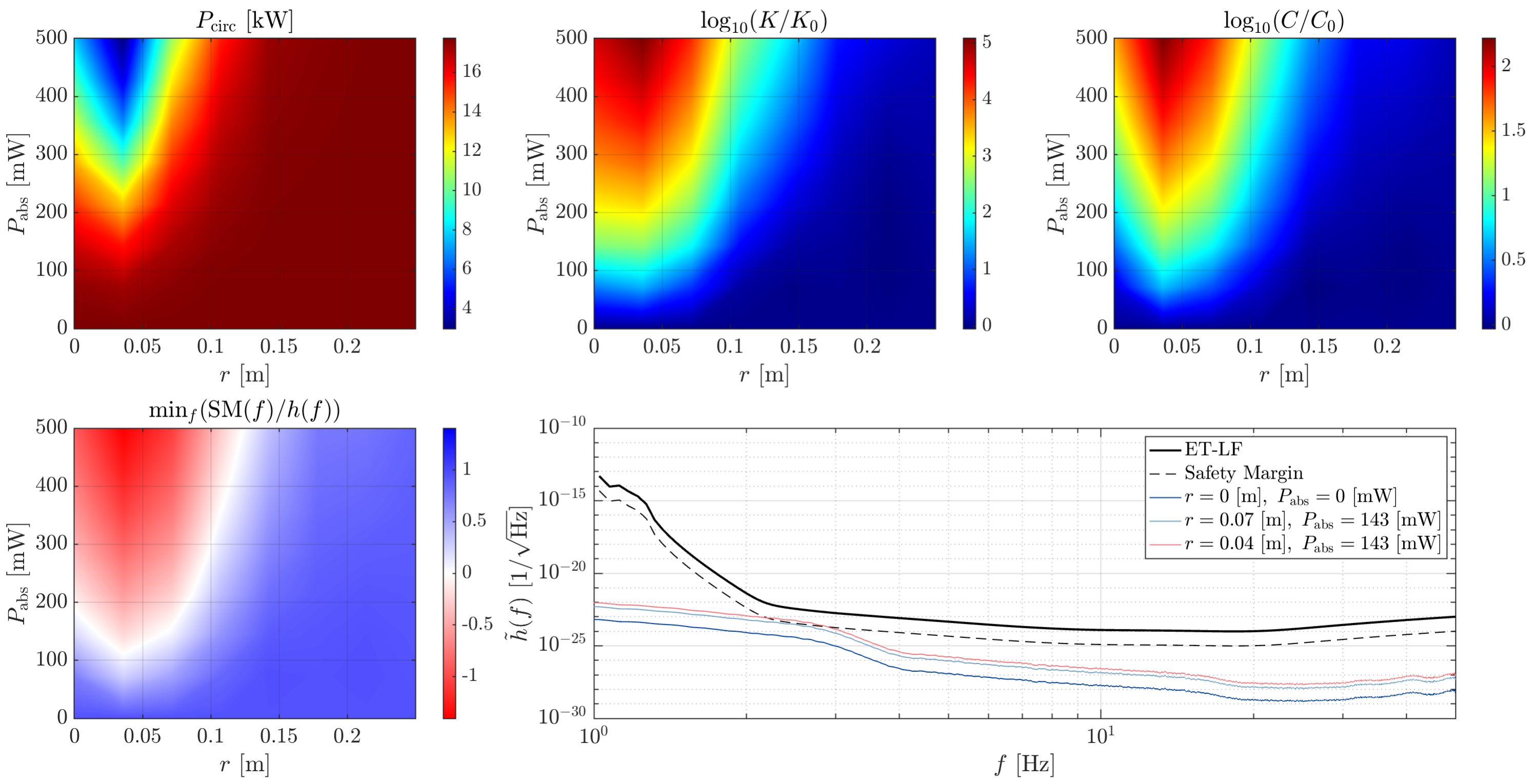}
     \caption{(top) For $\Delta$-shaped ET-LF, the circulating power in the cavity $P_{\rm circ}$ as a function of the position $\td r$ and the power absorption $P_{\mathrm{abs}}$ of the point defect in the mirror; and the evolution of backscattering, $K/K_0$, and diffraction, $C/C_0$, noise terms as a function of $\td r$ and $P_{\mathrm{abs}}$, where $K_0$ and $C_0$ denote the corresponding values for $\td r =  0, \ P_{\rm abs} = 0$.
    (bottom) For $\Delta$-shaped ET-LF, the resulting ${\rm{min}}_f (\mathrm{SM}(f)/h(f))$ as a function of $\td r$ and $P_{\rm abs}$,  and the resulting scattered light noise curve for different  $\td r$ and  $P_{\rm abs}$ values compared to the ET-LF sensitivity curve (black line) and the corresponding 1/10 safety margin (dashed line).}
    \label{fig:ETLFabs}
\end{figure}

Figures~\ref{fig:ETHFabs} and~\ref{fig:ETLFabs} collect the results for ET-HF and ET-LF, respectively.  As expected, the effect of the point absorber on the ET sensitivity strongly depends on its position in the mirror, the absorbed power, and the size of the laser beam, as computed at the mirror surface.  For ET-HF (ET-LF) defects located at $r>0.2$~m ($r > 0.15$~m) do not have a significant impact on the sensitivity. The scattered-light noise increases with increasing $P_{\rm{abs}}$.  As shown in the figures,  single point absorbers located at $r < 0.1$~m and with $P_{\rm{abs}} >100$~mW  would have the potential to compromise the ET sensitivity  in the range  $4-6$~Hz for ET-HF, and  in the range  $2-4$~Hz for ET-LF.  

\section{Conclusions}
\label{sec:sum}
We presented updated results on  the simulated 
stray-light noise in the main arms of  ET for both $\Delta$-shaped 10~km and L-shaped 15~km arm lengths and for high- and low-frequency configurations. We first consider the scenario of perfectly centered laser beams in the cavity. 
The simulations indicate that, with the considered baffle apertures and baffle layout along the cavity,  the stray-light noise remains below the safety margin and it does not constitute a limiting factor for the ET sensitivity. For the case of a $\Delta$-shaped 10~km arm, separate studies have been performed including scenarios with significant laser beam offsets, tilts, and the presence of defects in the form of point absorbers in the mirrors.  With the currently proposed baffle design,  offsets in the transverse plane, with respect to the laser beam direction, above 4~cm (7~cm) for the ET-HF (ET-LF) or tilts larger than about 8~$\mu$rad could already constitute a limitation for reaching the sensitivity of the experiment at low frequencies, in the range between 2~Hz and 6~Hz. To achieve higher tolerances, a larger baffle aperture and a redesign of the baffle layout along the cavity should be considered.  The presence of  point absorbers  in the mirrors is also a source of concern if they are located close to the center of the mirror (at distances in the transverse plane typically below 0.1~m) and with an absorption power exceeding about 100~mW. Altogether, these results indicate the range of parameters for beam misalignment and mirror defects that would be still compatible with reaching the expected ET sensitivity at given frequencies. 

The work does not include 
crucial considerations on the stray light noise levels originated 
in the vicinity of the main mirrors, inside the cryotrap areas attached to the vacuum towers hosting the mirror's suspensions, or inside the vacuum towers themselves. They are subject of an ongoing separate study.  

\clearpage
\section*{Acknowledgements}

This
project has received funding from the European Union’s Horizon 2020 research and innovation programme under the Marie Skłodowska-Curie Grant Agreement No. 754510.
This work
is partially supported by the Spanish MCIN/AEI/10.13039/501100011033 under the Grants No. PID2020-113701GB-I00 and PID2023-146517NB-I00
, some
of which include ERDF funds from the European Union, and by the MICINN with funding from the European Union NextGenerationEU (PRTR-C17.I1) and by the Generalitat de
Catalunya. IFAE is partially funded by the CERCA program of the Generalitat de Catalunya.
MAC was supported by the 2022 FI-00335 grant. This document has received an internal document number of ET-0284A-25. Part of this material is based upon work supported by NSF’s LIGO Laboratory which is a major facility fully funded by the National Science Foundation and operates under Cooperative Agreement No. PHY-1764464.

\bibliographystyle{iopart-num}
\bibliography{ref}

\end{document}